\documentclass[]{aastex}

\usepackage{emulateapj5}
\usepackage{onecolfloat}
\usepackage{graphicx}
\usepackage{fancyheadings}
\usepackage{ulem}
\usepackage{rotating}
\usepackage{lscape}

\def \cmm  {cm$^{-2}$}

\def \kms  {km~s$^{-1}$}
\def \lyaf {Ly$\alpha$ forest}

\def \lya  {Ly$\alpha$}

\def \ly5  {Ly-5}
\def \ly6  {Ly-6}
\def \ly7  {Ly-7}
\def \nhi  {$N_{\rm HI}$}
\def \mnhi  {N_{\rm HI}}
\def \lnhi {$\log N_{HI}$}
\def \mlnhi {\log N_{HI}}

\def \om {$\Omega_m$}
\def \ol {$\Omega_{\Lambda}$}
\def \gz {$g(z)$}
\def \mgz {g(z)}
\def \lyaf {Lyman--$\alpha$ forest}
\def \hfreq {$f_{\rm{HI}} (N,X)$}

\def \mllls {\ell_{\rm{LLS}}(X)}
\def \lslls {$\ell_{\rm{SLLS}}(X)$}
\def \loz {$\ell(z)$}
\def \lox {$\ell(X)$}
\def \ldla {$\ell_{\rm{DLA}}$}
\def \hi {\ion{H}{1}}
\def \O {${\mathcal O}(N,X)$}
\newcommand{\cm}[1]{\, {\rm cm^{#1}}}

\def \nslls {78}  

\begin{document}

\twocolumn[%
\submitted{Accepted to ApJ; Revised October 30 2006}

\title{The Keck $+$ Magellan Survey for Lyman Limit Absorption I:  
The Frequency Distribution of Super Lyman Limit Systems\altaffilmark{1}}

\author{
  John M. O'Meara\altaffilmark{2,3}, 
Jason X. Prochaska\altaffilmark{3,4},\\ 
Scott Burles\altaffilmark{5}, Gabriel Prochter\altaffilmark{3,4}, 
Rebecca A. Bernstein\altaffilmark{6}, and Kristin M. Burgess\altaffilmark{7}
\\}

\begin{abstract}
We present the results of a survey for super Lyman limit systems
(SLLS; defined to be absorbers with $19.0 \le $\lnhi\ $\le 20.3$ \cmm)
from a large sample of high resolution spectra acquired using the 
Keck and Magellan telescopes.  Specifically, we present 47 new SLLS 
from 113 QSO sightlines.  We focus on the neutral hydrogen frequency distribution 
\hfreq\ of the SLLS and its moments, and compare these results
with the \lyaf\ and the damped Lyman alpha systems (DLA;
absorbers with \lnhi$\ge 20.3$ \cmm). 
We find that that \hfreq\ of the SLLS can be reasonably described with a power-law of index $\alpha = -1.43^{+0.15}_{-0.16}$ or 
$\alpha = -1.19^{+0.20}_{-0.21}$ depending on whether we set the 
lower \nhi\ bound for the analysis at $10^{19.0}$\cmm\ or $10^{19.3}$\cmm,
respectively.  
The results indicate a flattening in the 
slope of \hfreq\ between the SLLS and DLA.
We find little evidence for redshift evolution in 
the shape of \hfreq\ for the SLLS 
over the redshift range of the sample $1.68 < z < 4.47$
and only tentative evidence for evolution in the zeroth moment 
of \hfreq, the line density \lslls.  
We introduce the observable distribution function \O\ 
and its moment, which 
elucidates comparisons of H~I absorbers from the \lyaf\
through to the DLA.  We find that a simple three parameter function can fit \O\ over the range $17.0 \le $\lnhi$\le 22.0$.  
We use these results to  
predict that \hfreq\ must show two additional inflections below the SLLS regime 
to match the observed \hfreq\ distribution of the \lyaf. 
Finally, we demonstrate that SLLS contribute a minor fraction
$(\approx 15\%$) of the universe's hydrogen atoms and, therefore,
an even smaller fraction of the mass in predominantly neutral gas.

\keywords{quasars:  absorption lines -- intergalactic medium}
\end{abstract}
]

\altaffiltext{1}{This paper includes data gathered with the 6.5 meter Magellan Telescopes located at Las Campanas Observatory, Chile.}
\altaffiltext{2}{Department of Physics, Penn State Worthington Scranton, 120 Ridge View Drive, Dunmore, PA 18512}
\altaffiltext{3}{Department of Astronomy and Astrophysics, UCO/Lick Observatory, University of California, 1156 High Street, Santa Cruz, CA 95064}
\altaffiltext{4}{Visiting Astronomer, W.M. Keck Observatory which is a joint facility of the University of California, the California Institute of Technology, and NASA}
\altaffiltext{5}{MIT Kavli Institute for Astrophysics and Space Research,Massachusetts Institute of Technology, 77 Massachusetts Avenue, Cambridge MA 02139}
\altaffiltext{6}{Department of Astronomy, University of Michigan, Ann Arbor, MI 48109}
\altaffiltext{7}{Department of Physics, Princeton University, Princeton, NJ 08544}

\pagestyle{fancyplain}
\lhead[\fancyplain{}{\thepage}]{\fancyplain{}{O'MEARA ET AL.}}
\rhead[\fancyplain{}{Super Lyman Limit Systems}]{\fancyplain{}{\thepage}}
\setlength{\headrulewidth=0pt}
\cfoot{}

\section{Introduction}
For nearly three decades, the study of absorption line systems 
towards distant quasars has addressed a wide range of astrophysical and 
cosmological issues.  These systems are 
typically classified according to their neutral hydrogen content: the \lya\ forest absorbers with \lnhi $\le 17.2$ \cmm, the Lyman limit systems (LLS) with $17.2 \le $\lnhi $\le 20.3$ \cmm, and the damped Lyman alpha systems (DLA) with \lnhi $\ge 20.3$ \cmm.  Both the \lya\ forest and DLA absorbers have received considerable attention.  The \lya\ forest can be used to constrain cosmological parameters through a number of methods, such as through studies of the flux power spectrum 
\citep{cwb+03,mcdonald05}, 
the mean flux decrement \citep{tytler04}, 
or the distribution in column density and velocity width of the absorbers 
\citep[e.g.][]{kt97,kim02}. 
The DLA trace the bulk of the neutral gas at high redshift and 
are believed to be the progenitors of modern 
day galaxies \citep{wgp05}.  Large statistical samples 
of both the \lya\ forest and the DLA are readily observed 
because their \nhi\ is easily determined.  For the \lya\ forest, the \nhi\ is determined  directly from Voigt profile fits to the \lya\ line which is
dominated by the Maxwellian profile, or from higher order Lyman transitions when \lya\ is saturated. 
For the DLA, the Lorentzian component of the Voigt profile 
gives pronounced damping wings, allowing for accurate \nhi\ determinations from spectra even at low signal to noise or resolution.  

Recently, Prochaska, Herbert-Fort, \& Wolfe (2005, hereafter PHW2005) analyzed the thousands of spectra from the Sloan Digital Sky Survey (SDSS) Data Release 3, and determined the \nhi\ frequency
distribution for over 500 DLA systems.
In contrast, there exists comparatively little study of the LLS. 
Surveys for LLS absorption \citep{tytler82,ssb,lzt91,storrie94,key_lls}
have concentrated primarily on the 
frequency of absorption with redshift (frequently expressed as
$dn/dz$ or $dN/dz$, but we adopt the notation \loz\ for the line
density which was introduced by PHW2005), but not 
on the \nhi\ value of the LLS.  These surveys generally included 
the full range of \ion{H}{1} column density of 
$\mlnhi \ge 17.2$ \cmm\ in \loz. As such, the surveys also contained 
 the \loz\ of DLA systems. 

Our ignorance of LLS largely stems from the fact that 
accurate determinations of \nhi\ for the LLS are difficult compared to the \lya\ forest and the DLA.  
In part, this is because the LLS systems represent the flat portion
of the curve of growth for the \lya\ transition and a precise \nhi\ measurement
requires high resolution observations \citep[e.g.][]{steidel90}.
For most LLS, the \nhi\ is determined by using the information from the Lyman limit, either by looking at the differential flux level above and below the limit, or by the spacing of the lines as they approach the Lyman limit, or both
\citep[e.g.][]{llsfit}.  
The differential flux method requires a detailed understanding of the continuum flux level which is difficult to obtain in high redshift QSOs, and the line spacing technique requires a precise model for the hydrogen velocity structure, which can generally only be inferred from the metal lines associated with the LLS \citep[e.g.][]{llsvel}.
These challenges are particularly acute for low-resolution observations
where the Lyman series is poorly resolved.

Because the frequency of intersecting a LLS is observed to 
be of order one per sightline at redshifts $z>2.5$ \citep{storrie94}, a relatively
large QSO sample is required in comparison with \lya\ forest studies.  
At redshifts $z > 3.25$, the SDSS provides just such a sample \citep{sdssdr4}, 
offering many thousands of sightlines suitable for analysis of LLS.  Unfortunately, 
the SDSS spectra are of too poor a resolution to provide useful constraints 
on the \nhi\ for the LLS which are optically thick, i.e. \lnhi $\ge 17.5$ \cmm.  
To address this nearly three decade gap in \nhi, 
we have pursued a high-resolution survey for 
LLS using telescopes in both the northern and southern hemisphere.  The goal of this survey is to obtain high--resolution, high--SNR spectra of at least 100 LLS, with full coverage over the LLS \nhi\ range.  In this paper, we discuss a sub--sample of this survey, namely the LLS with \lnhi $\ge 19.0$ \cmm.  The lower bound on \nhi\ is chosen such that systems with this column density 
can be easily identified and analyzed in moderate to high resolution 
data \citep[FWHM~$< 50$ \kms;][]{mirka03}.  
The \nhi\ values for these LLS are determined from the damping wings present in the \lya\ line, in a fashion analogous with DLA analysis in lower resolution
spectra.
Previous work on this subset of the LLS population referred to the 
absorption systems as `sub-DLAs'.  Because the majority of these
absorbers are likely to be 
predominantly ionized \citep[e.g.][]{viegas95,prochaska99}, 
we adopt the nomenclature of \cite{ph04}
for those LLS exhibiting \lnhi $\ge 19.0$ \cmm:  
the `super Lyman limit' or SLLS absorbers. 

The fundamental measure of a class of QAL systems is the \ion{H}{1}
column density frequency distribution \hfreq, defined to be the number of
absorbers in the column density interval $(N,N+dN)$ identified
along the cosmological distance interval $(X,X+dX)$ with
$dX \equiv \frac{H_0}{H(z)} (1+z)^2 dz$. 
This quantity, the absorption distance,
is generally evaluated across a redshift interval
$\Delta z$.  The \hfreq\ distribution for QAL systems is analogous
to the luminosity function of galaxy surveys.  Moments of \hfreq\
give important quantities such as the line density of absorption systems, and the mass density
of \ion{H}{1} atoms. The \hfreq\ distribution is the starting point for assessing
the baryonic mass density of these absorbers as well as their
cosmological metal budget.
The frequency distribution of LLS is of particular interest because
the $\mnhi = 10^{17} - 10^{20} \cm{-2}$  interval is expected to
include the \nhi\ value where QAL systems transition from primarily
neutral gas to predominantly ionized gas 
\citep[e.g.][; PHW05]{zheng02}.
Furthermore, \cite{poh+06} have 
argued that the LLS can constitute a considerable fraction of 
the metals in the young universe \citep[see also][]{pkm06}.
At $z \simeq 2$, a census for metals which includes the DLA, 
stars in high--$z$ galaxies, 
and the IGM falls short by up to 70\% of the 
total predicted metal mass density \citep[e.g.][]{bouche06II}.

The primary goal for this study of the SLLS is to extend the statistics we use to describe the \lyaf\ and the DLA into regions of H~I column density which are currently poorly constrained \citep[see also][]{peroux_slls03,peroux05}.  
By doing so, we hope to place the SLLS within the larger framework of high redshift QSO absorption line systems, and to explore their cosmological significance.
Future papers will examine the ionization state, chemical abundances,
and other physical properties of these absorbers.
Throughout the paper, we adopt values of the cosmological parameters consistent
with the latest  \textit{Wilkinson Microwave Anisotropy Probe (WMAP)} 
results \citep{wmap03}: \ol $=0.7$, \om$ = 0.3$, and $H_0 = 70$ \kms\ Mpc$^{-1}$.

\begin{table*}\footnotesize 
\begin{center}
\caption{{\sc MIKE QUASAR SAMPLE\label{qsotab_MIKE}}}
\begin{tabular}{lcccccl}
\tableline
\tableline
Name &RA (J2000) & DEC (J2000) &$z_{em}$ & $z_{start}$ & $z_{end}$ &
$z_{mask}^a$ \\
\tableline
  Q0001-2340   &00:03:45.00&$-$23:23:46.5&2.262&1.780&2.228&2.187\\
  Q0101-304    &01:03:55.30&$-$30:09:46.0&3.137&1.941&3.095&\\
  SDSS0106+0048&01:06:19.24&$+$00:48:23.3&4.433&2.997&4.378&\\
  SDSS0124+0044&01:24:14.80&$+$00:45:36.2&3.807&2.292&3.758&\\
  SDSS0147-1014&01:45:16.59&$-$09:45:17.3&2.138&1.797&2.106&\\
  SDSS0209-0005&02:09:50.70&$-$00:05:06.4&2.856&1.879&2.816&2.523\\
  SDSS0244-0816&02:44:47.78&$-$08:16:06.1&4.047&2.829&3.996&\\
  HE0340-2612  &03:42:27.80&$-$26:02:43.0&3.082&2.016&3.040&\\
  SDSS0912+0547&09:12:10.35&$+$05:47:42.0&3.248&2.146&3.205&\\
  SDSS0942+0422&09:42:02.04&$+$04:22:44.6&3.273&1.896&3.229&\\
  HE0940-1050  &09:42:53.40&$-$11:04:25.0&3.067&1.944&3.025&\\
  SDSS0949+0355&09:49:32.27&$+$03:35:31.7&4.097&2.636&4.045&\\
  SDSS1025+0452&10:25:09.64&$+$04:52:46.7&3.236&2.108&3.193&\\
  SDSS1032+0541&10:32:49.88&$+$05:41:18.3&2.843&1.826&2.804&\\
  CTS0291      &10:33:59.90&$-$25:14:26.7&2.552&1.747&2.515&\\
  SDSS1034+0358&10:34:56.31&$+$03:58:59.3&3.367&2.208&3.322&\\
  Q1100-264    &11:03:25.60&$-$26:45:06.1&2.140&1.747&2.108&\\
  HS1104+0452  &11:07:08.40&$+$04:36:18.0&2.660&1.747&2.622&\\
  SDSS1110+0244&11:10:08.61&$+$02:44:58.1&4.149&3.367&4.097&\\
  SDSS1155+0530&11:55:38.60&$+$05:30:50.5&3.464&2.282&3.418&\\
  SDSS1201+0116&12:01:44.36&$+$01:16:11.5&3.215&2.002&3.172&\\
  LB1213+0922  &12:15:39.60&$+$09:06:08.0&2.713&1.796&2.675&\\
  Q1224-0812   &12:26:37.50&$-$08:29:29.0&2.142&1.747&2.110&\\
  SDSS1249-0159&12:49:57.24&$-$01:59:28.8&3.662&2.406&3.614&\\
  SDSS1307+0422&13:07:56.73&$+$04:22:15.5&3.026&1.821&2.985&\\
  LBQS1334-0033&13:36:46.80&$-$00:48:54.2&2.809&1.829&2.770&\\
  SDSS1336-0048&13:36:47.14&$-$00:48:57.2&2.806&1.796&2.767&\\
  SDSS1339+0548&13:39:41.95&$+$05:48:22.1&2.969&1.969&2.928&\\
  HE1347-2457  &13:50:38.90&$-$25:12:17.0&2.599&1.749&2.562&\\
  Q1358+1154   &14:00:39.10&$+$11:20:22.3&2.578&1.747&2.541&\\
  SDSS1402+0146&14:02:48.07&$+$01:46:34.1&4.187&2.618&4.134&\\
  SDSS1429-0145&14:29:03.03&$-$01:45:19.3&3.416&2.331&3.371&\\
  Q1456-1938   &14:56:49.83&$-$19:38:52.0&3.163&1.879&3.120&\\
  SDSS1503+0419&15:03:28.88&$+$04:19:49.0&3.666&3.126&3.618&\\
  SDSS1521-0048&15:21:19.68&$-$00:48:18.6&2.935&2.178&2.895&\\
  SDSS1558-0031&15:58:10.15&$-$00:31:20.0&2.831&1.784&2.792&\\
  Q1559+0853   &16:02:22.60&$+$08:45:36.3&2.267&1.747&2.218&1.842,2.251\\
  SDSS1621-0042&16:21:16.92&$-$00:42:50.8&3.704&2.142&3.656&\\
  Q1720+2501   &17:22:52.90&$+$24:58:34.7&2.250&1.961&2.217&\\
  PKS2000-330  &20:03:24.10&$-$32:51:44.0&3.776&2.422&3.727&\\
  Q2044-1650   &20:47:19.70&$-$16:39:05.8&1.939&1.747&1.909&\\
  Q2053-3546   &20:53:44.60&$-$35:46:52.4&3.484&2.154&3.438&\\
  SDSS2100-0641&21:00:25.03&$-$06:41:46.0&3.118&2.093&3.076&\\
  SDSS2123-0050&21:23:29.46&$-$00:50:52.9&2.278&1.837&2.244&2.059\\
  Q2126-158    &21:29:12.20&$-$15:38:41.0&3.278&1.943&3.234&\\
  Q2147-0825   &21:49:48.20&$-$08:11:16.2&2.127&1.879&2.095&\\
  SDSS2159-0021&21:59:54.45&$-$00:21:50.1&1.963&1.797&1.932&\\
  HE2156-4020  &21:59:54.70&$-$40:05:50.0&2.530&1.747&2.494&\\
  HE2215-6206  &22:18:51.00&$-$61:50:43.0&3.317&1.720&3.273&\\
  Q2249-5037   &22:52:44.00&$-$50:21:37.0&2.870&1.788&2.830&\\
  SDSS2303-0939&23:03:01.45&$-$09:39:30.7&3.453&2.241&3.407&\\
  HE2314-3405  &23:16:43.20&$-$33:49:12.0&2.944&1.684&2.904&\\
  SDSS2346-0016&23:46:25.67&$-$00:16:00.4&3.467&2.166&3.421&\\
  HE2348-1444  &23:51:29.80&$-$14:27:57.0&2.933&1.837&2.893&2.279\\
  HE2355-5457  &23:58:33.40&$-$54:40:42.0&2.931&1.854&2.891&\\
\tableline
\end{tabular}
\end{center}
\tablenotetext{a}{These redshifts correspond to LLS or DLA
which were known to exist along the QSO sightline prior to 
the higher resolution observations and also which inspired the 
observations.}
\end{table*}

\section{Spectroscopic Sample}
The quasar sample in this paper includes spectra from two instruments, the Magellan Inamori Kyocera Echelle  \citep[MIKE;][]{bernstein03} high 
resolution spectrograph on the Magellan 6.5m telescope at Las Campanas Observatory in Chile, and the Echellete Spectrograph and Imager 
\citep[ESI;][]{sheinis02} on the Keck-II 10m telescope in 
Hawaii. MIKE is a double echelle spectrograph, with a dichroic optical element splitting the beam into blue and red arms, each with their own CCD. MIKE provides full wavelength coverage, without spectral gaps from 3350--9500 \AA\ in the default configuration. When a $1.0''$ slit is used, MIKE has $R=28,000$ and $R=22,000$ for the blue and red sides, respectively. ESI is a
spectrograph and imager, which provides continuous 
wavelength coverage from 3900--10900 \AA\ in echellette mode.
When a $0.5''$ slit is used, ESI provides $R \approx 9,000$.

In table \ref{qsotab_MIKE}, we list the 57 QSOs in the current MIKE sample.  
For each QSO in the sample, the data was reduced using the MIKE
reduction pipeline\footnote{http://www.lco.cl/lco/magellan/instruments/MIKE/index.html} 
\citep{bbp06}.
The pipeline flat-fields, optimally extracts, flux calibrates, and combines exposures to produce a single spectrum for the red and blue CCDs of MIKE.  
In Table \ref{qsotab_ESI}, we list the 56 QSOs in the ESI sample.  The bulk of these spectra come from the survey for DLA absorption presented by 
\cite{pro03} but supplemented by a new sample of SDSS spectra.
All of the ESI data were reduced with the ESIRedux 
pipeline\footnote{http://www2.keck.hawaii.edu/inst/esi/ESIRedux/index.html}
\citep{pro03}.
The difference in native resolution of the two data-sets will have implications
for the \nhi\ completeness limit of each survey.  This is discussed
in greater detail in the following sections.

\begin{table*}\footnotesize
\begin{center}
\caption{{\sc ESI QUASAR SAMPLE\label{qsotab_ESI}}}
\begin{tabular}{lcccccl}
\tableline
\tableline
Name & RA (J2000) & DEC (J2000) &$z_{em}$ &$z_{start}$ & $z_{end}$ &
$z_{mask}^a$ \\
\tableline
  SDSS0013+1358&00:13:28.21&$+$13:58:27.0&3.565&2.747&3.518&3.281\\
  PX0034+16    &00:34:54.80&$+$16:39:20.0&4.290&3.031&4.207&4.260\\
  SDSS0058+0115&00:58:14.31&$+$01:15:30.3&2.535&2.373&2.499&\\
  SDSS0127-00  &01:27:00.70&$-$00:45:59.0&4.066&2.907&4.014&3.727\\
  PSS0134+3307 &01:34:21.60&$+$33:07:56.0&4.525&3.154&4.469&3.761\\
  SDSS0139-0824&01:39:01.40&$-$08:24:43.0&3.008&2.373&2.967&2.677\\
  SDSS0142+0023&01:42:14.74&$+$00:23:24.0&3.363&2.356&3.303&3.347\\
  SDSS0225+0054&02:25:54.85&$+$00:54:51.0&2.963&2.331&2.922&2.714\\
  SDSS0316+0040&03:16:09.84&$+$00:40:43.2&2.907&2.331&2.866&\\
  BRJ0426-2202 &04:26:10.30&$-$22:02:17.0&4.328&3.039&4.274&2.980\\
  FJ0747+2739  &07:47:11.10&$+$27:39:04.0&4.119&2.767&4.066&3.900,3.423\\
  PSS0808+52   &08:08:49.40&$+$52:15:15.0&4.440&3.155&4.385&3.113,2.942\\
  SDSS0810+4603&08:10:54.90&$+$46:03:58.0&4.072&3.442&4.020&2.955\\
  FJ0812+32    &08:12:40.70&$+$32:08:09.0&2.700&2.290&2.662&2.626\\
  SDSS0816+4823&08:16:18.99&$+$48:23:28.4&3.578&2.784&3.531&3.437\\
  Q0821+31     &08:21:07.60&$+$31:07:35.0&2.610&2.348&2.573&2.535\\
  SDSS0826+5152&08:26:38.59&$+$51:52:33.2&2.930&2.331&2.795&2.834,2.862\\
  SDSS0844+5153&08:44:07.29&$+$51:53:11.0&3.193&2.373&3.150&2.775\\
  SDSS0912+5621&09:12:47.59&$-$00:47:17.4&2.967&2.373&2.851&2.890\\
  Q0930+28     &09:33:37.30&$+$28:45:32.0&3.436&2.529&3.391&3.246\\
  PC0953+47    &09:56:25.20&$+$47:34:42.0&4.463&3.154&4.407&4.245,3.889,3.403\\
  PSS0957+33   &09:57:44.50&$+$33:08:23.0&4.212&2.981&4.124&4.177,3.280\\
  SDSS1004+0018&10:04:28.43&$+$00:18:25.6&3.042&2.422&3.001&2.540\\
  BQ1021+30    &10:21:56.50&$+$30:01:41.0&3.119&2.290&3.076&2.949\\
  CTQ460       &10:39:09.50&$-$23:13:26.0&3.134&2.290&3.091&2.778\\
  HS1132+22    &11:35:08.10&$+$22:27:15.0&2.879&2.290&2.839&2.783\\
  BRI1144-07   &11:46:35.60&$-$07:40:05.0&4.153&2.916&4.101&\\
  PSS1159+13   &11:59:06.48&$+$13:37:37.7&4.071&2.751&4.019&3.724\\
  Q1209+09     &12:11:34.90&$+$09:02:21.0&3.271&2.743&3.227&2.586\\
  PSS1248+31   &12:48:20.20&$+$31:10:43.0&4.308&3.031&4.254&3.698\\
  PSS1253-02   &12:53:36.30&$-$02:28:08.0&3.999&2.948&3.948&2.782\\
  SDSS1257-0111&12:57:59.22&$-$01:11:30.2&4.100&2.414&3.972&4.022\\
  Q1337+11     &13:40:02.60&$+$11:06:30.0&2.915&2.373&2.875&2.796\\
  PSS1432+39   &14:32:24.80&$+$39:40:24.0&4.276&3.014&4.222&3.272\\
  HS1437+30    &14:39:12.30&$+$29:54:49.0&2.991&2.290&2.950&2.874\\
  SDSS1447+5824&14:47:52.47&$+$58:24:20.2&2.971&2.389&2.930&2.818\\
  SDSS1453+0023&14:53:29.53&$+$00:23:57.5&2.531&2.373&2.495&2.444\\
  SDSS1610+4724&16:10:09.42&$+$47:24:44.5&3.201&2.373&3.158&2.508\\
  PSS1723+2243 &17:23:23.20&$+$22:43:58.0&4.515&3.006&4.459&3.695\\
  SDSS2036-0553&20:36:42.29&$-$05:52:60.0&2.575&2.414&2.538&2.280\\
  FJ2129+00    &21:29:16.60&$+$00:37:56.6&2.954&2.290&2.913&2.735\\
  SDSS2151-0707&21:51:17.00&$-$07:07:53.0&2.516&2.406&2.480&2.327\\
  SDSS2222-0946&22:22:56.11&$-$09:46:36.2&2.882&2.784&2.842&2.354\\
  Q2223+20     &22:25:36.90&$+$20:40:15.0&3.574&2.344&3.527&3.119\\
  SDSS2238+0016&22:38:43.56&$+$00:16:47.0&3.425&2.455&3.321&3.365\\
  PSS2241+1352 &22:41:47.70&$+$13:52:03.0&4.441&3.483&4.385&4.283\\
  SDSS2315+1456&23:15:43.56&$+$14:56:06.0&3.370&2.373&3.326&3.273\\
  PSS2323+2758 &23:23:40.90&$+$27:57:60.0&4.131&2.907&4.078&3.684\\
  FJ2334-09    &23:34:46.40&$-$09:08:12.0&3.326&2.307&3.282&3.057\\
  SDSS2343+1410&23:43:52.62&$+$14:10:14.0&2.907&2.373&2.867&2.677\\
  Q2342+34     &23:44:51.20&$+$34:33:49.0&3.030&2.735&2.989&2.908\\
  SDSS2350-00  &23:50:57.87&$-$00:52:09.9&3.010&2.866&2.969&2.615\\
\tableline
\end{tabular}
\end{center}
\tablenotetext{a}{These redshifts correspond to LLS or DLA
which were known to exist along the QSO sightline prior to 
the higher resolution observations and also which inspired the 
observations.}
\end{table*}

\subsection{UVES Sample}
In addition to the MIKE and ESI data, we include the results of the surveys by 
\cite{peroux_slls03,peroux05}, 
which we will refer to as the ``UVES sample''.  
All of these data were drawn from a heterogeneous sample of high resolution
observations using the UVES spectrometer \citep{uves} on the VLT-2 telescope.
In five cases, there is an 
overlap in quasars observed between our sample and the UVES sample.  
In these cases, we remove those quasars which contribute the smaller redshift path to the total sample.  

\subsection{Redshift Path}
For each quasar in our sample, we define a redshift interval 
$\Delta z = z_{end} - z_{start}$ to construct the redshift path 
in a manner similar to that presented by PHW2005.  The starting
redshift $z_{start}$ is given by the lowest redshift at which we could identify strong \lya\ features. For the majority of sightlines, this  is set by
the starting wavelength of the spectrum $\lambda_0$: 
$z_{start} = \lambda_0/1215.67{\rm \AA} - 1$.  Higher values for $z_{start}$ were adopted for a sightline when there was either intervening LLS absorption which completely removes the QSO flux, or when the SNR of the QSO became so low as to significantly impact the likelihood of detecting a strong \lya\ feature.
The ending redshift $z_{end}$ is given by $z_{end} = 0.99z_{em} - 0.01$, where
$z_{em}$ corresponds to the QSO redshift.  
Thus, $z_{end}$ is defined to be located 
$\approx 3000$ \kms\ blueward of the QSO \lya\ emission line.  
This offset was chosen to remove those LLS which could be 
associated with the QSO.  Tables \ref{qsotab_MIKE} and \ref{qsotab_ESI} list $z_{start}$ and $z_{end}$ for each QSO in our sample. Also given 
in Tables \ref{qsotab_MIKE} and \ref{qsotab_ESI} are the redshifts, $z_{DLA}$, of any DLA known to be present in the data prior
to the observations.  It is our expectation that SLLS are strongly
clustered with DLA \citep[e.g.][]{pw99} and we wish to avoid biasing the sample 
(because these systems frequently motivated the observations).
We mask out regions 1500 \kms\ on either side of the DLA redshifts 
from the redshift path (corresponding to $\approx$15\,comoving Mpc $h^{-1}$ at $z=3$)
to prevent SLLS clustering with DLA from biasing the sample.
In this fashion we construct a sensitivity function $g_i(z)$   
which has unit value at redshifts where SLLS could be detected and zero otherwise.
The combined redshift path of our sample is $\Delta z = 124.4$.

The \ion{H}{1} frequency distribution function 
\hfreq\ for the SLLS describes 
the number of SLLS in a range of column densities $(N, N + dN)$, 
and a range of absorption distances $(X, X+ dX)$,

\begin{equation}
f_{\rm{HI}} (N,X)dNdX, 
\end{equation}
where the absorption distance \citep{bp69} is given by

\begin{equation}
\Delta X = \int dX = \int \frac{H_0}{H(z)}(1+z)^2 dz .
\end{equation}
By considering $dX$ instead of $dz$, \hfreq\ is defined over a constant,
comoving pathlength but we have introduced our assumed cosmology
into the analysis \citep[e.g.][]{lzt93}.
The cosmological term in \hfreq\ is still 
relatively obscure, but can be made less so when we consider that for
$z >2$, $\frac{dX}{dz} \approx \sqrt{(1 + z)/\Omega_{m}}$ for a 
standard $\Lambda$CDM model.  
By way of example, consider a sightline with $\Delta z=1$ pathlength
at the average redshift 
for our survey $\bar z=3.1$ corresponding to $\Delta X \approx 3.7$. 

\begin{figure}
\includegraphics[height=3.5in,angle=90]{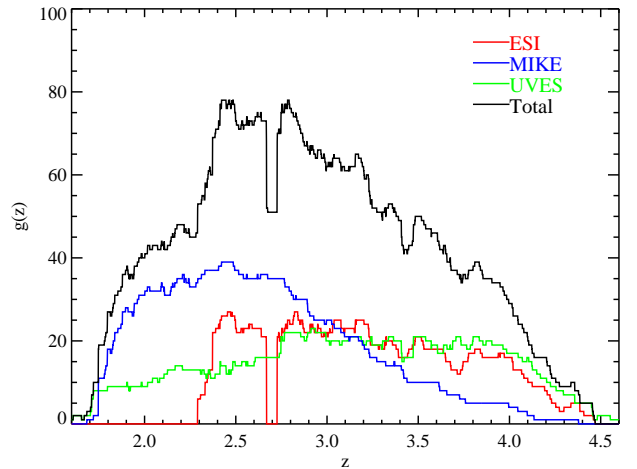}
\caption{Redshift sensitivity function \gz\ for the samples.  The sharp feature at 
$z \simeq 2.7$ is caused by a gap in the ESI data.  The upper (black) curve represents
 the full sample \gz subject to the added constraint that it apply to the data where a SLLS with \nhi\ $>19.3$ \cmm could be found, with the other curves (blue, red, green) representing the
 \gz\ for the MIKE, ESI, and UVES data samples which comprise the full sample.  
}
\label{fig_gz}
\end{figure}

To characterize the survey size, we construct the total 
redshift sensitivity function, $\mgz = \Sigma_i g_i(z)$.
Figure \ref{fig_gz} shows the \gz\ curves
for the MIKE, ESI and UVES samples, and the combined \gz\ for all three samples.   We find that the total sample presented here increases \gz\ by 
factors of 2 to 8 
over the redshift range $2 \le z \le 4$ when compared to 
the previous survey of \cite{peroux05} when we consider the column density range 19.3 \cmm $\le $\lnhi\ $\le 20.2$ \cmm. 
In Figure \ref{fig_gz}, there is a sharp feature in \gz\ at a redshift of $z \simeq 2.7$.  This feature is due to a gap in the ESI data at $\lambda \simeq 4500$\AA\ which appears in every ESI spectrum in the sample.  At the 
redshifts surveyed, the resolution 
of the data is sufficiently high that sky lines do not effect 
our ability to detect LLS in the data.

\section{\nhi\ Analysis}
With the survey size defined, we now turn to the SLLS in the survey and discuss the measurement of the \nhi\ for the SLLS, and an estimate of the completeness of the sample.

\begin{figure*}
\includegraphics[width=6.8in]{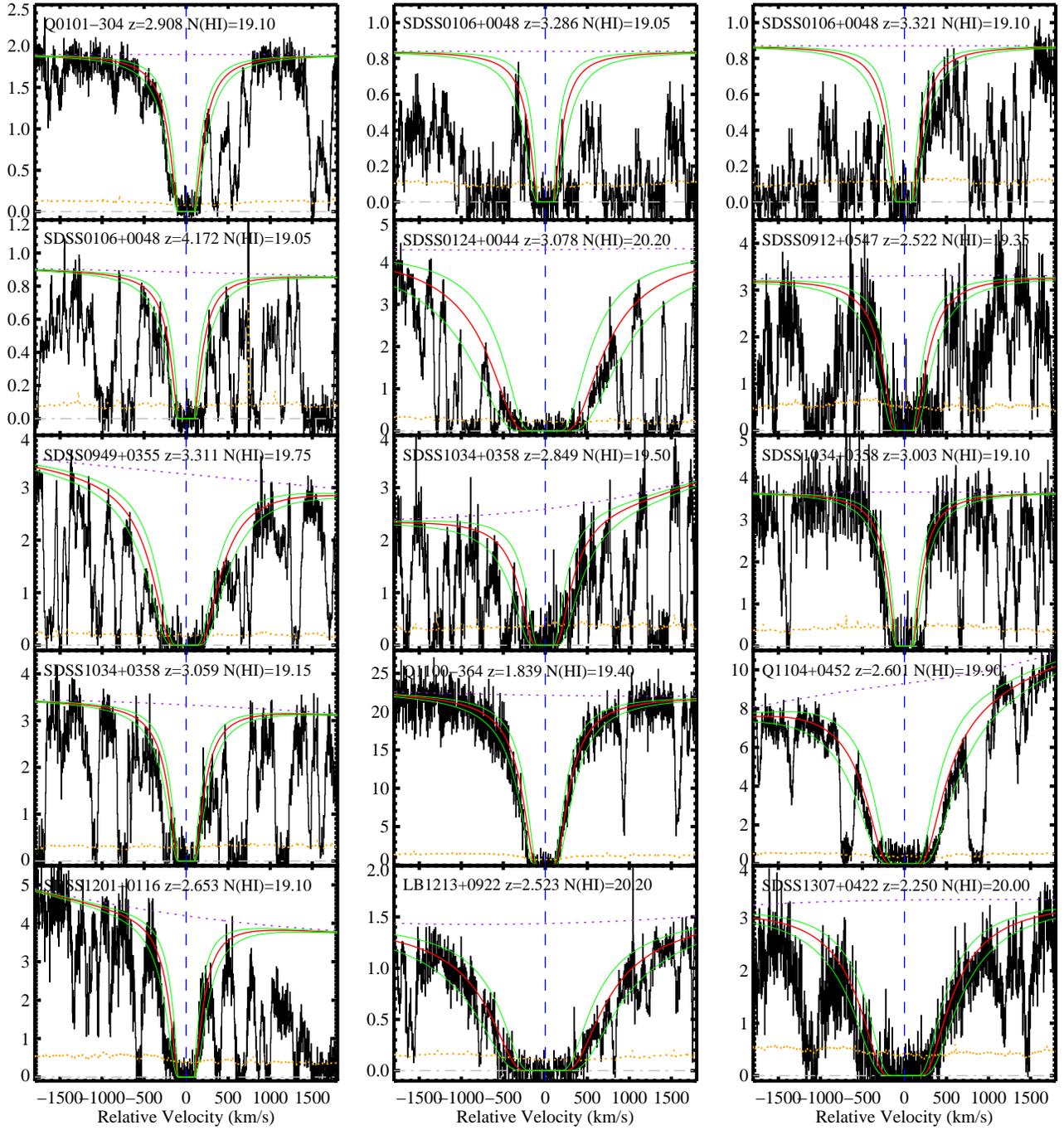}
\caption{
H~I Lyman--$\alpha$ transitions for the SLLS in the MIKE sample. The  velocity zeropoint is determined from low ionization metal lines whenever possible.  Overlaid (in red) are the single component Voigt profile models for the absorption along with 
the $\pm 1\sigma$ error estimates (in green)
on the \nhi.  The horizontal dashed line represents the local continuum level. 
 The vertical dashed line highlights the velocity zeropoint.  For each absorber
, the QSO name, absorption redshift, and \nhi\ estimate are listed at the top of
 each panel.  In the case of blended SLLS where the individual \lya\ lines are separated by more than 300 \kms, we show the individual fits
 (e.g. the SLLS in PKS2000-330).}
\label{fig_mikefits}
\end{figure*}

\begin{figure*}
\epsscale{0.8}
\includegraphics[width=6.8in]{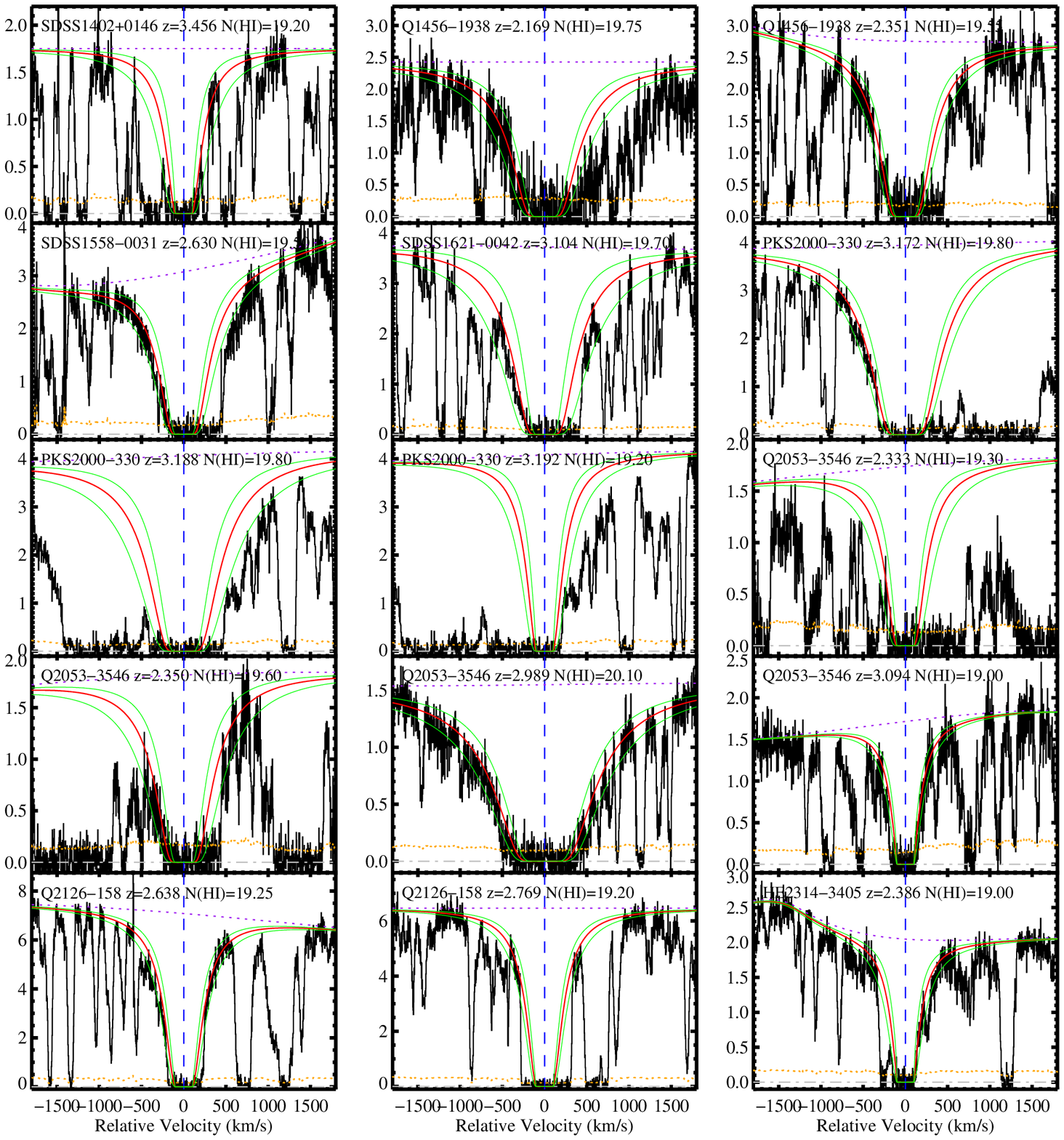}
\newline
\center{Fig. 2 -- continued}
\end{figure*}

\subsection{Identifying SLLS and Measuring \nhi}

To identify the SLLS in our survey, every spectrum from the ESI and MIKE samples was 
visually inspected for strong absorption in \ion{H}{1} \lya.  
When a candidate SLLS absorber is identified, we first assign a local continuum level to regions approximately 2000 \kms\ 
on either side of the absorption.  We do not first continuum normalize the data, opting instead to model the absorption and continuum 
simultaneously.  This is done because the damping wings of the 
\lya\ profile may depress the continuum for several tens of
Angstroms.  It is also necessary to 
vary the continuum level along with a model to
estimate the errors on the \nhi\ values.  
When the absorption occurs near the QSO emission 
line features and/or when the redshift of the absorption is 
high,  the assignment of the continuum level is often 
the dominant source of error.  In this work, 
we assign a \textit{minimum} value of the error on 
the \lnhi\ to be 0.05~dex from  continuum level errors alone.

Next, we compared a Voigt profile with \lnhi\ $=19.0$ \cmm\ at the center 
of the absorption to determine if it was consistent with at 
least this amount of atomic hydrogen gas.
If metal-line transitions were identified, we use the redshift of the metals to more accurately assign a redshift for the gas.  
We do not, however, require that metal line absorption be present, 
because this would introduce an undesirable 
bias.  For most of the absorption systems in our sample, the data cover a wide variety of metal line transitions for each absorber.  Common transitions include those of C~II, C~IV, O~I, Si~II, Si~IV, Al~II, and Fe~II.   In the cases where both low and high ionization metal line transitions are present, we choose to use the redshift given by the low-ion absorption, because these ions more likely trace the atomic hydrogen gas.  
Finally, for those absorbers meeting our minimum \nhi\ condition, we 
fit the absorption by stepping through values of \nhi\ until the 
absorption is well modeled. The \lya\ line profile and continuum 
level were modeled using custom software
(the \textit{x\_fitdla} tool within the XIDL 
package\footnote{http://www.ucolick.org/$\sim$xavier/IDL/}). 
In the cases when two or more
SLLS occur within $\delta v \le 300$ \kms\ of each other, the 
\textit{total} \nhi\ is reported instead of 
individual \nhi\ measurements.  This is because 
the individual \nhi\ measurements are highly degenerate and 
because the gas may physically bound to a single virialized halo.

For each absorber, a number of effects contribute to the error in the \nhi\
value.  As mentioned above, we believe that errors in the continuum level contribute at least 0.05 dex to the error.  The continuum level error tends to increase 
with redshift because the amount of absorption from \lyaf\ increases and 
one is less certain to identify regions free of absorption. 
The increase in \lyaf\ lines also affects the \nhi\ measurement in
the damping wings and line core regions of the \lya\ line because
of enhanced line-blending. 
Redshift uncertainties were only a  minor source of error for each absorber, since nearly every absorption system showed at least one metal-line transition. 
Finally, the Poisson noise of the data adds additional uncertainty to
the fitted value.

\begin{figure*}
\includegraphics[width=6.8in]{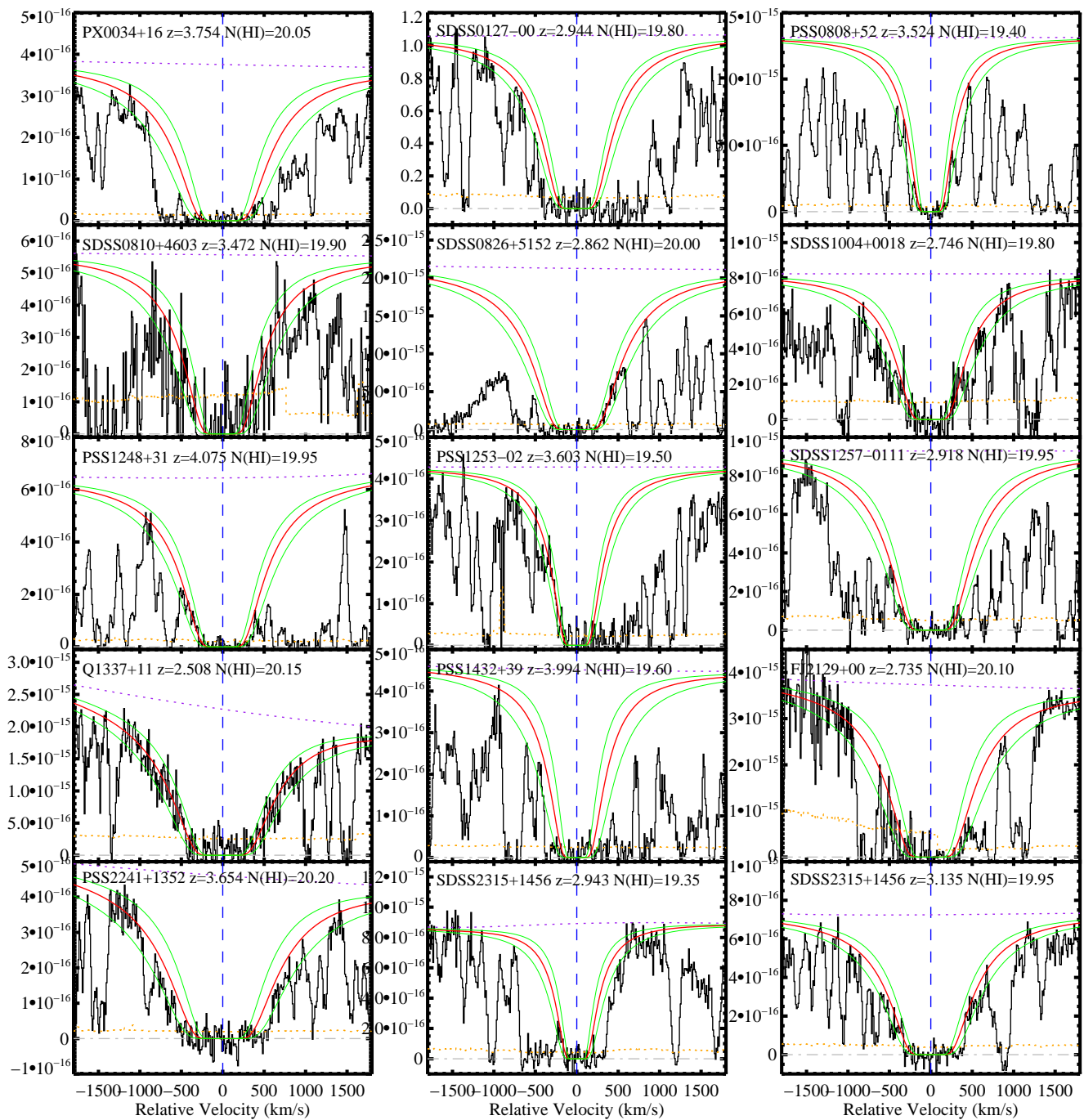}
\caption{
Same as for Figure \ref{fig_mikefits}, but for the SLLS in the ESI sample.
}
\label{fig_esifits}
\end{figure*}

\begin{figure}
\epsscale{0.8}
\includegraphics[width=3.5in, trim= 0 100 0 0 ]{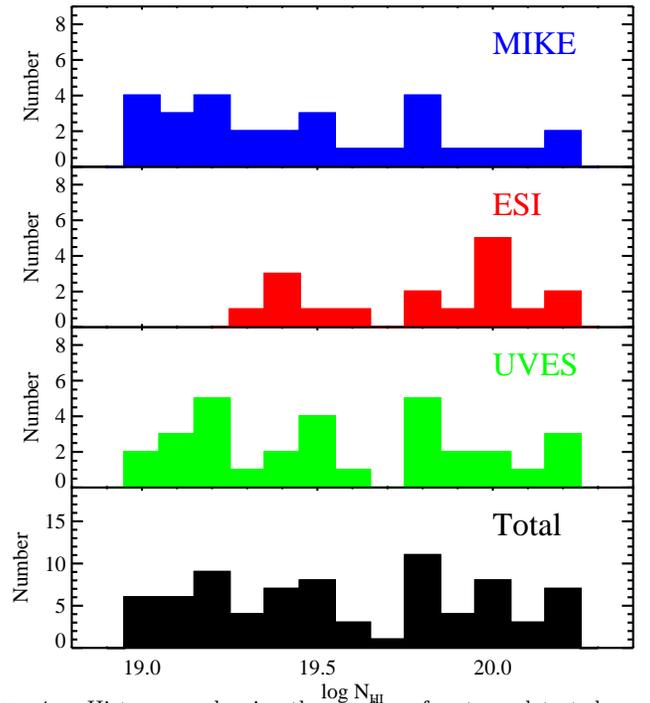}
\caption{
Histograms showing the number of systems detected as a function of \lnhi\ for the various samples.  The upper two panels 
represent the MIKE and ESI data, which are presented for the first time in this 
paper.  The third panel shows the data from \cite{peroux05}. The bottom 
panel shows the full \nhi\
histogram when we combine the three data sets.
}
\label{fig_hihist}
\end{figure}

In Figures \ref{fig_mikefits} and \ref{fig_esifits}, we show the profile fits and the $\pm 1 \sigma$ fits for each LLS in our sample.  The values 
for the $z$, \nhi, and $\sigma(N_{\rm HI})$ are given in Tables \ref{tab_mikenhi} and \ref{tab_esinhi}.  The LLS listed in these tables do not represent every LLS present in our data.  
In a number of cases, 
we removed an SLLS from the sample because the QSO was specifically targeted 
to study this SLLS. For those SLLS removed, we also remove 1500 \kms\ of pathlength on either side of the SLLS in the same manner as for the DLA.
In total, we have a homogeneous sample of 47 SLLS, of which 17 come from the ESI sample and 30 come from the MIKE sample.  Including 
the UVES sample, there are now a total of \nslls\ SLLS. 
In Figure \ref{fig_hihist}, we show a histogram of the \nhi\ values for the individual and complete samples.  

\begin{table}[ht]\footnotesize
\begin{center}
\caption{{\sc LLS FROM THE MIKE SAMPLE\label{tab_mikenhi}}}
\begin{tabular}{lccc}
\tableline
\tableline
Name &$z_{abs}$ & $\log N_{\rm HI}$ & $\sigma(N_{\rm HI})$ \\
\tableline
Q0101-304&2.908&19.10&0.15\\
SDSS0106+0048&3.286&19.05&0.25\\
SDSS0106+0048&3.321&19.10&0.20\\
SDSS0106+0048&4.172&19.05&0.20\\
SDSS0124+0044&3.078&20.20&0.20\\
SDSS0912+0547&2.522&19.35&0.20\\
SDSS0949+0355&3.311&19.75&0.15\\
SDSS1034+0358&2.849&19.50&0.25\\
SDSS1034+0358&3.059&19.15&0.15\\
SDSS1034+0358&3.003&19.10&0.15\\
Q1100-264&1.839&19.40&0.15\\
HS1104+0452&2.601&19.90&0.20\\
LB1213+0922&2.523&20.20&0.20\\
SDSS1307+0422&2.250&20.00&0.15\\
SDSS1402+0146&3.456&19.20&0.30\\
Q1456-1938&2.351&19.55&0.15\\
Q1456-1938&2.169&19.75&0.20\\
SDSS1558-0031&2.630&19.50&0.20\\
SDSS1621-0042&3.104&19.70&0.20\\
PKS2000-330&3.188&19.80&0.25\\
PKS2000-330&3.172&19.80&0.15\\
PKS2000-330&3.192&19.20&0.25\\
Q2053-3546&2.350&19.60&0.25\\
Q2053-3546&2.989&20.10&0.15\\
Q2053-3546&3.094&19.00&0.15\\
Q2053-3546&2.333&19.30&0.25\\
Q2126-158&2.638&19.25&0.15\\
Q2126-158&2.769&19.20&0.15\\
HE2314-3405&2.386&19.00&0.20\\
\tableline
\end{tabular}
\end{center}
\end{table}

\begin{table}[ht]\footnotesize
\begin{center}
\caption{{\sc LLS FROM THE ESI SAMPLE\label{tab_esinhi}}}
\begin{tabular}{lccc}
\tableline
\tableline
Name &$z_{abs}$ & $\log N_{\rm HI}$ & $\sigma(N_{\rm HI})$ \\
\tableline
PX0034+16&3.754&20.05&0.20\\
SDSS0127-00&2.944&19.80&0.15\\
PSS0808+52&3.524&19.40&0.20\\
SDSS0810+4603&3.472&19.90&0.20\\
SDSS0826+5152&2.862&20.00&0.15\\
SDSS1004+0018&2.746&19.80&0.20\\
PSS1248+31&4.075&19.95&0.15\\
PSS1253-02&3.603&19.50&0.15\\
SDSS1257-0111&2.918&19.95&0.15\\
Q1337+11&2.508&20.15&0.15\\
PSS1432+39&3.994&19.60&0.25\\
FJ2129+00&2.735&20.10&0.20\\
PSS2241+1352&3.654&20.20&0.20\\
SDSS2315+1456&3.135&19.95&0.15\\
SDSS2315+1456&2.943&19.35&0.20\\
PSS2323+2758&3.267&19.40&0.20\\
PSS2323+2758&3.565&19.30&0.20\\
\tableline
\end{tabular}
\end{center}
\end{table}

\subsection{Completeness}
To gauge the completeness of our \nhi\ analysis we performed a number of tests.  Our primary completeness concern is with the detection efficiency at the low H~I column density limit of the ESI sample, where the data has lower spectral resolution.  
At lower resolution, the effects of line blending and continuum level placement tend to wash out the damping wings of \lnhi $ \simeq 19.0 $ \lya\ lines.  As the column density increases, the equivalent width of the \lya\ line also increases
and one generally derives a more  
reliable \nhi\ measurement.  Because 
the MIKE and UVES samples are of higher resolution, the effect 
at lower column densities is minimized.  
After experimenting with mock ESI spectra (see below), we chose to 
limit the ESI analysis to those 
absorbers with 19.3 \cmm $\le $ \lnhi $\le 20.3$ \cmm.  For the MIKE sample, we consider the same range of column densities as in the UVES sample, 19.0 \cmm $\le $ \lnhi\ $\le 20.3$ \cmm.  Our tests indicate
that we have detected all SLLS in these ranges in \nhi\ at a 99\% level of completeness.

\begin{figure}[ht]
\includegraphics[scale=0.65, trim= 0 500 0 0]{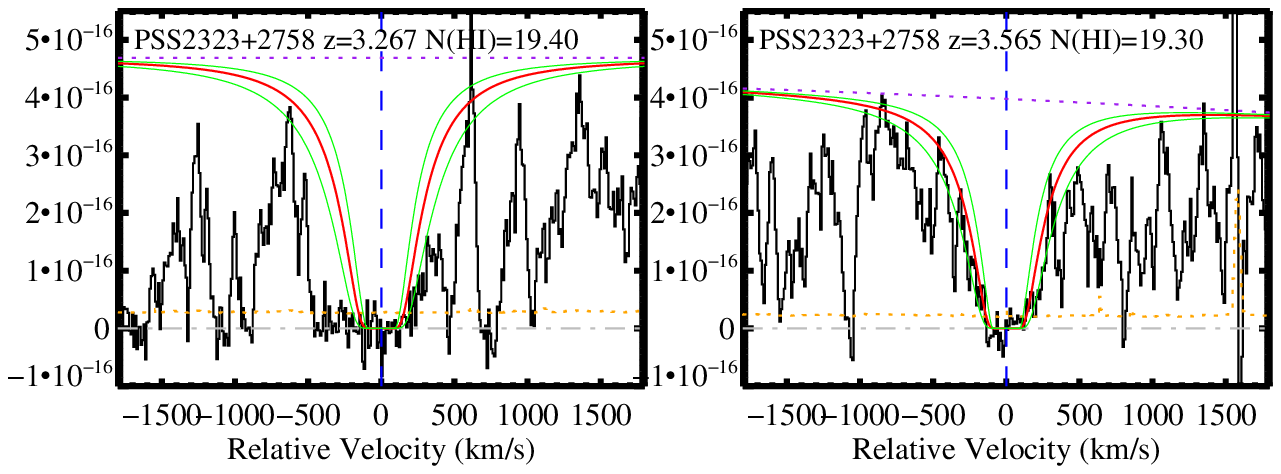}
Fig. 3 -- continued
\end{figure}

To verify that the lower bound of \lnhi $=19.3$ \cmm\ is appropriate for 
the ESI sample, we randomly placed mock LLS onto 50 of the 
ESI spectra. The column densities of these LLS were chosen at random to lie 
in the range $18.0 \le $\lnhi $\le 20.5$ \cmm.  In all cases, SLLS with \lnhi $ \ge 19.3$ \cmm\ were identified.  Furthermore, the column density fit to these SLLS were within the 2--$\sigma$ error estimates in all cases but one out of fifty.  We have also used simulated LLS to address the issue that blending of multiple LLS with \lnhi\ $\le 19.3$ \cmm\  could appear as a single absorber with \lnhi\ $\ge 19.3$ \cmm\ in the ESI sample.  As before, we placed mock LLS upon ESI spectra, but in pairs with a random velocity separation 
$\delta v$ subject to the constraint $\delta v \le 300$ \kms.  
This constraint matches our velocity separation constraint for the observed SLLS as to whether the \nhi\ is reported as a sum, or as separate systems. 
The blended, mock 
LLS absorption was then fit with a single absorption profile, and the \nhi\ compared to the sum of the individual component \nhi\ values.  We find that a pair of LLS with \lnhi $\le 18.8 \cm{-2}$ or less can not blend in a manner so as to be fit as an SLLS with \lnhi\ $\ge 19.3$ \cmm.  
However, a blend of a SLLS with \lnhi\ $\ge 19.3$ \cmm with a LLS with \lnhi\ $\sim 19.1$ \cmm\ generally leads to a systematic overestimate of \nhi.  Similarly, a blend of LLS with \lnhi $ = 19.1$ \cmm\ and \lnhi\ = $18.8$\cmm\ may mimic an SLLS with \lnhi\ $=19.3$ \cmm. 
In essence, this effect leads to a Malmquist bias for the sample at
low \nhi\ value.
It is not generally possible to directly explore these final two issues without good knowledge of the incidence frequency, clustering properties, and column density distribution of the LLS.  Nevertheless, we have some constraints from both the MIKE and UVES samples.  Because these samples are of higher resolution, we can often distinguish blended absorption of LLS from a single LLS.  Moreover, if two LLS are blended, their metal lines should give some clues that there is blending, provided both LLS have sufficient metallicity. To the extent that we can currently distinguish between blended and unblended LLS, we do not believe that the effect of blended LLS strongly effect our measurements of the \nhi\, but we caution that a larger sample than the one
presented here is required to accurately address this issue.

\section{Results and Discussion}

In this section we discuss a number of results related to the
\ion{H}{1} frequency distribution of the SLLS.
For the majority of the section, we consider two \nhi\ groups
driven by the completeness limits of the spectra:
(i) $10^{19.0} \cm{-2} \leq \mnhi \leq 10^{20.3} \cm{-2}$ 
which does not include the
ESI sample; and 
(ii) $10^{19.3} \cm{-2} \leq \mnhi \leq 10^{20.3} \cm{-2}$ 
which includes all of the samples.
The groups cover an integrated absorption 
pathlength of $\Delta X^{19.0} = 329.1$  
and $\Delta X^{19.3} = 467.7$ respectively.   
We maintain this division, as opposed to combining the 
low \nhi\ results from the echelle data with the 
$\mnhi \geq 10^{19.3} \cm{-2}$
analysis, for the following reasons:  
First, the redshift distribution of the ESI survey
(Figure~\ref{fig_gz}) is significantly higher than that of the UVES data
and especially the MIKE sample.
If there is redshift evolution in \hfreq, then it would
be erroneous to mix the two groups.  
Indeed, PHW05 find significant evolution int he normalization of
\hfreq\ for the damped \lya\ systems and one may expect similar
evolution in the SLLS population.
Second, the division allows us to investigate systematic
errors between the various surveys, including the UVES analysis.
Finally, we will find it instructive
valuable to consider these two \nhi\ groups separately
when focusing on the behavior of \hfreq\ near $\mnhi = 10^{19} \cm{-2}$.

\subsection{Power-Law Fits to the \hfreq\ Distribution of the SLLS}

\begin{figure}
\epsscale{0.8}
\includegraphics[width=3.5in, trim= 0 100 0 0 ]{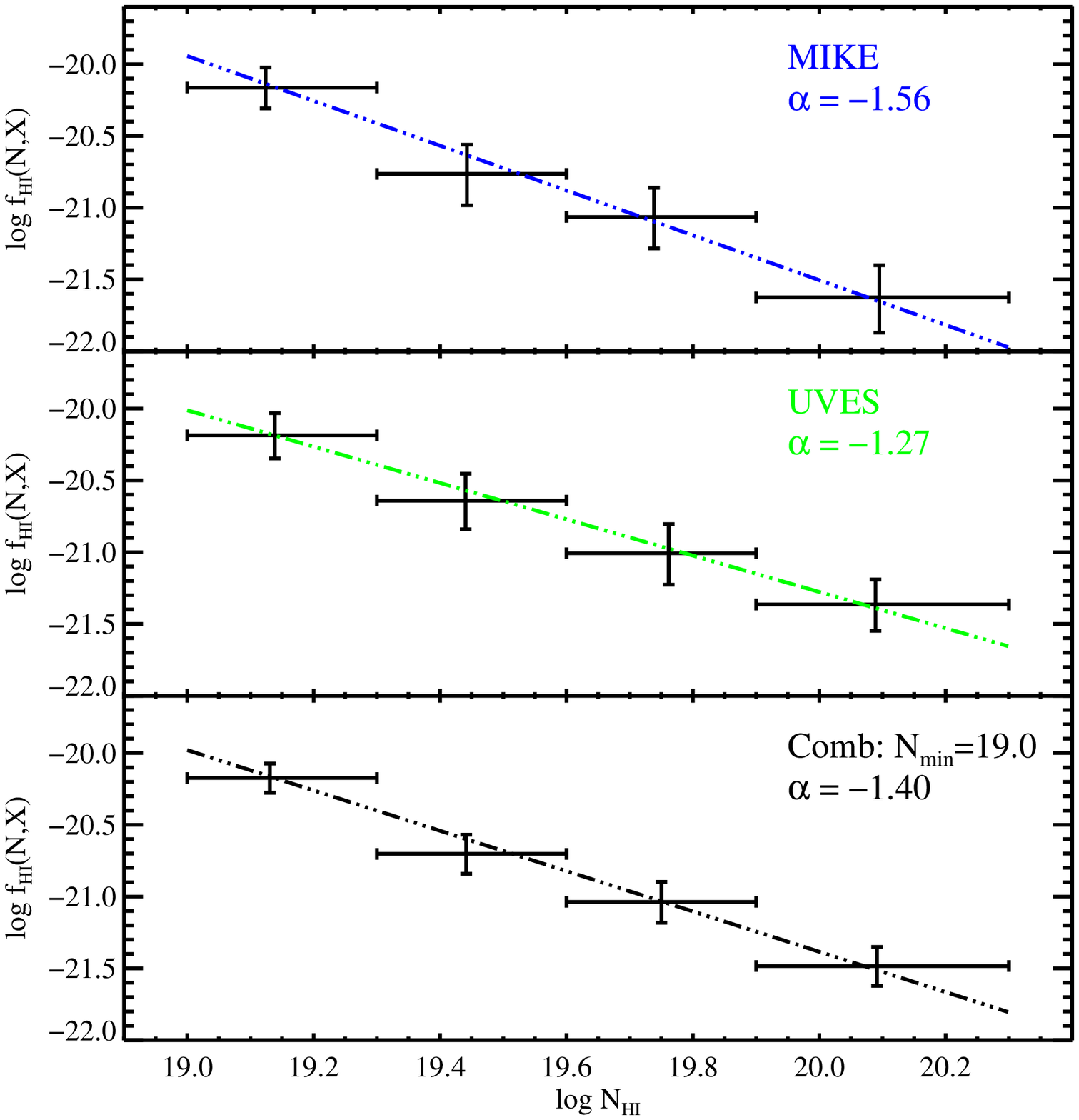}
\caption{
Values for the H~I frequency distribution for the SLLS with $19.0 \le$ \lnhi\ $20.3$\cmm.  Shown separately are the MIKE data presented first here, the UVES sample of \cite{peroux_slls03,peroux05}, and the two samples combined. The four bins span equal size $\Delta$\lnhi$=0.3$ except for the last bin which has $\Delta$\lnhi$=0.4$.  
Also shown is the result of a maximum likelihood analysis on the unbinned data for a power-law model of \hfreq.  The power-law index $\alpha$ is given for each sample.}
\label{fig_fnall190}
\end{figure}

\begin{figure}
\epsscale{0.8}
\includegraphics[width=3.5in, trim= 0 75 0 0 ]{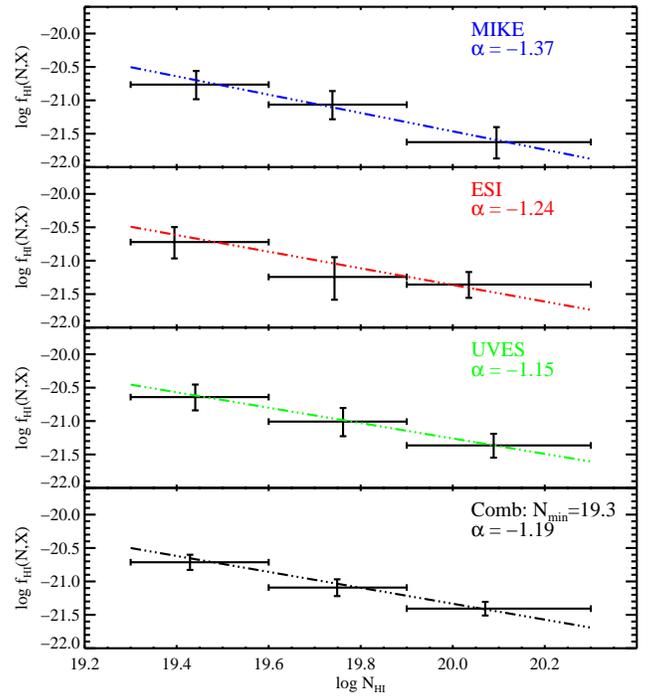}
\caption{Same as for Figure \ref{fig_fnall190}, but with the inclusion of the ESI data, and with the range $19.0 \le$ \lnhi\ $20.3$ \cmm, over 3 bins equal in size to the 3 highest H~I column density bins in Figure \ref{fig_fnall190}.
Also shown is the result of a maximum likelihood analysis on the unbinned data for a power-law model of \hfreq.  The power-law index $\alpha$ is given for each sample.
}
\label{fig_fnall193}
\end{figure}

In Figures \ref{fig_fnall190} and \ref{fig_fnall193}, we show the binned
evaluation of \hfreq\ for each of the data-sets, and their combined results.  
The results are summarized in Table \ref{tab_fn}.
It is evident from the figures that the \hfreq\ distributions can be 
reasonably modeled by a power-law over the SLLS regime:

\begin{equation}
f_{\rm{HI}}\left(N,X\right) = k_1 N^{\alpha} \;\; .
\end{equation}
We have performed a maximum likelihood analysis on the unbinned
\nhi\ distribution to constrain $\alpha$ and set the normalization of
\hfreq.  We find $\alpha_{190} = -1.40^{+0.15}_{-0.16}$ 
and $\alpha_{193} =-1.19^{+0.20}_{-0.21}$ for the combined results
of the two groups.  
Note that the reported errors do not include covariance terms.
The results for the individual data
samples are all consistent (within $2\sigma$) with the combined
results and one another (Table~\ref{tab_fn}).

\begin{table*}[ht]\footnotesize
\begin{center}
\caption{\hfreq\ SUMMARY\label{tab_fn}}
\begin{tabular}{lcccccccccc}
\tableline
\tableline
Sample & $\Delta X$ & $\bar z$ & $m$ & & & log\hfreq 
& & $\ell_{\rm SLLS}(X)$ & $\log k_1$ & $\alpha_1$ \\
& & & & $N \epsilon [19.0, 19.3)$ & 
$N \epsilon [19.3, 19.6)$ & 
$N \epsilon [19.6, 19.9)$ & 
$N \epsilon [19.9, 20.3)$ \\
\tableline
ESI&132.5&3.34& 15&...&$-20.72^{+0.22}_{-0.24}$&$-21.24^{+0.30}_{-0.34}$&$-21.36^{+0.19}_{-0.20}$&$0.113^{+0.037}_{-0.029}$&$ 3.527^{+0.124}_{-0.128}$&$-1.244^{+0.390}_{-0.404}$\\
MIKE&175.6&2.85& 29&$-20.16^{+0.14}_{-0.15}$&$-20.76^{+0.20}_{-0.22}$&$-21.06^{+0.20}_{-0.22}$&$-21.63^{+0.22}_{-0.24}$&$0.165^{+0.037}_{-0.030}$&$ 9.735^{+0.087}_{-0.088}$&$-1.562^{+0.224}_{-0.240}$\\
UVES&154.2&3.33& 31&$-20.19^{+0.15}_{-0.16}$&$-20.64^{+0.19}_{-0.20}$&$-21.01^{+0.20}_{-0.22}$&$-21.36^{+0.17}_{-0.18}$&$0.201^{+0.043}_{-0.036}$&$ 4.033^{+0.084}_{-0.085}$&$-1.265^{+0.209}_{-0.216}$\\
ALL-19.3&467.7&3.08& 55&...&$-20.71^{+0.11}_{-0.12}$&$-21.09^{+0.12}_{-0.13}$&$-21.41^{+0.10}_{-0.10}$&$0.118^{+0.018}_{-0.016}$&$ 2.486^{+0.062}_{-0.062}$&$-1.191^{+0.203}_{-0.206}$\\
ALL-19.0&329.8&3.10& 60&$-20.17^{+0.10}_{-0.10}$&$-20.70^{+0.13}_{-0.14}$&$-21.04^{+0.14}_{-0.15}$&$-21.48^{+0.13}_{-0.14}$&$0.182^{+0.027}_{-0.023}$&$ 6.716^{+0.059}_{-0.059}$&$-1.405^{+0.153}_{-0.158}$\\
\tableline
\end{tabular}
\end{center}
\end{table*}

One notes that the power-law is shallower for
the $\mnhi^{19.3}$ sample than the $\mnhi^{19.0}$ sample.
Although the effect is not statistically significant, we
do find the trend is apparent in the independent MIKE and UVES
samples.  These results suggest that the \hfreq\ distribution
is steepening at $\mnhi < 10^{19.3} \cm{-2}$.
We will return to this point in $\S$~\ref{sec:inflect}.

\subsection{Redshift Evolution}

Over the redshift interval $z=2.2$ to 3.5, PHW2005 found that the 
shape of \hfreq\ for DLA is roughly constant but that the
normalization (parameterized by the zeroth moment, \ldla)
increases by a factor of approximately two.
Similarly, studies of the \lya\ forest indicate 
$\ell(z)^{Ly\alpha} \propto (1+z)^{2.6}$ \citep{cristiani95} at $z \approx 3$
implying 
$\ell(X)^{Ly\alpha} \propto (1+z)^{2.1}$.  
One may expect, therefore,
that \lslls\ will also increase with redshift over this redshift interval.

\begin{figure}
\epsscale{0.8}
\includegraphics[width=3.5in, trim= 0 150 0 0 ]{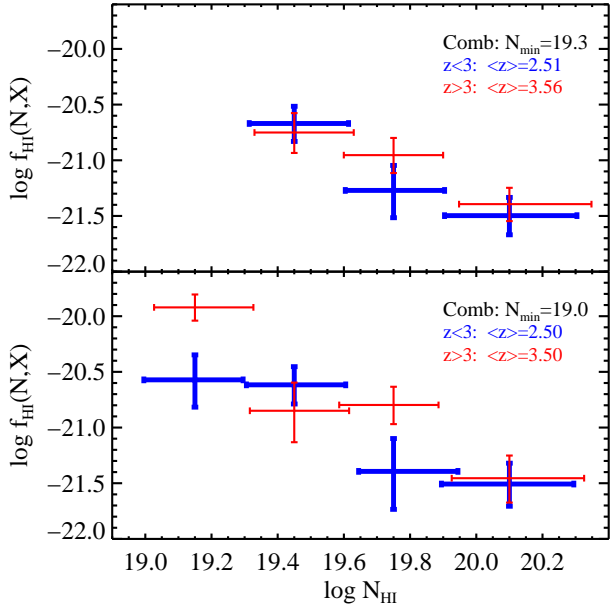}
\caption{Values for the H~I frequency distribution for the SLLS separated into two redshift samples, with the thin symbols corresponding to $z < 3$ and the thick symbols corresponding to $z > 3$.  The upper panel shows the ESI+MIKE+UVES sample and the lower panel the MIKE+UVES sample.  The lack of significant change in shape of \hfreq\ for the two redshift bins indicates a lack of strong evolution with redshift in \hfreq.
}
\label{fig_fnz}
\end{figure}

In Figure \ref{fig_fnz}, we
show the results for \hfreq\ in the two \nhi\ groups, 
but broken into the redshift
regions above (thin line) and below (thick line) $z=3$.  A visual inspection of Figure \ref{fig_fnz} 
reveals little explicit difference in the shape of 
\hfreq\ in the two redshift regimes.  
More formally, a KS test for the two groups returns probabilities 
of $P_{190} = 0.189$ and  $P_{193} = 0.809$ that the two redshift samples 
are drawn from the same parent population.  
Therefore, we contend there is no large redshift evolution in the shape
of \hfreq\ for the SLLS over the redshifts considered here.

Granted that the shape of \hfreq\ for the SLLS is not steeply
evolving, we can examine redshift variations in the normalization by
examining the zeroth moment

\begin{equation}
\ell_{\rm{SLLS}}(X) = \int\limits_{N_{SLLS}}^{10^{20.3}} f_{\rm{HI}} (N,X) dN
\end{equation}
with $N_{SLLS} = 10^{19.0}$ or $10^{19.3} \cm{-2}$.
We calculate \\
\lslls$_{(190)} = 0.134^{+0.032}_{-0.026}$ and  
\lslls$_{(193)}=0.110^{+0.026}_{-0.021}$ for $z < 3$ and 
\lslls$_{(190)} = 0.254^{+0.050}_{-0.042}$ and  
\lslls$_{(193)}=0.126^{+0.028}_{-0.023}$ for $z > 3$.  
As is evident from Figure~\ref{fig_fnz} (i.e.\ focus on the values for
the lowest \nhi\ bins), there is no significant
redshift evolution in \lslls\ for the $\mnhi = 10^{19.3} \cm{-2}$ sample.
There is, however, an indication of increasing \lslls\
with increasing redshift for the $\mnhi = 10^{19.0} \cm{-2}$ sample.
This result is driven by the lowest \nhi\ bin which we caution is the
most sensitive to the effects of line-blending and that such effects
are heightened at higher redshift.
In summary, there is only tentative evidence 
for redshift evolution in \lslls.
As noted above, this contradicts the apparent evolution in \lox\ for
the DLA population and the \lya\ forest.  If confirmed by future studies,
perhaps modest evolution in the SLLS population suggests the competing
effects of the decrease in the mean density of the universe and a
decrease in the intensity of the extragalactic UV background radiation field.

\subsection{Is \hfreq\ for the SLLS flatter than the DLA?}
\label{sec:flatten}

There is a significant mismatch in power-law exponents for the \lya\ forest 
($\alpha_{Ly\alpha} = -1.5$) and the DLA ($\alpha_{DLA} \approx  -2$)
at $z \approx 3$.
This difference predicts that the shape of \hfreq\ change 
appreciably between these two column densities, i.e.,   
the \hfreq\ distribution of SLLS is flatter than that of the
DLA.

The simplest test of this prediction is to measure the power-law
index of \hfreq\ in the SLLS regime and compare against its slope
near $\mnhi = 10^{20.3} \cm{-2}$.  Both the Gamma function and
double power-law fits to \hfreq\ for the SDSS DLA sample
indicate $\alpha_{DLA} = -1.9 \pm 0.1$ at $\mnhi < 10^{21} \cm{-2}$
(PHW2005).  We compare these values with 
$\alpha_{SLLS} = -1.19 \pm 0.21$ for $\log \mnhi^{lim} = 19.3$ \cmm and 
$\alpha_{SLLS} = -1.40 \pm 0.15$ for $\log \mnhi^{lim} = 19.0$ \cmm.
The statistical difference in the power-law slope is greater than
$2 \sigma$ significance but not a full $3\sigma$ result, in
terms of the exponents alone.

\begin{figure}
\epsscale{0.8}
\includegraphics[width=3.5in, trim= 0 150 0 0 ]{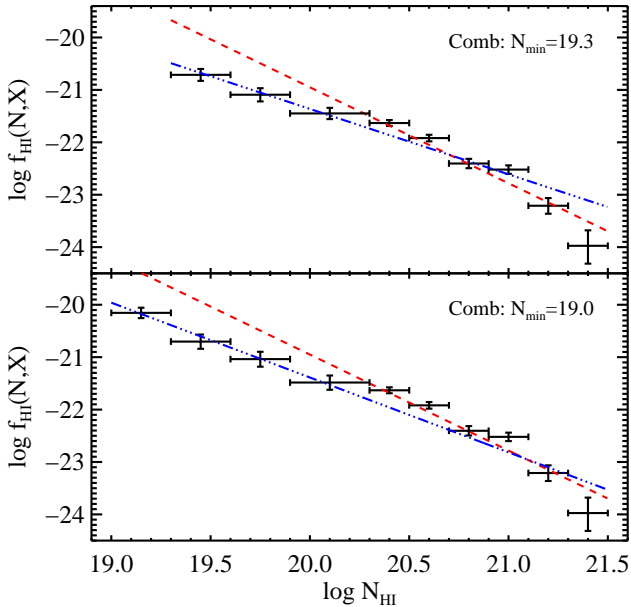}
\caption{Values for the H~I frequency distribution for both the SLLS and the DLA sample of PHW2005.  Overplotted are the results for the single power-law for the SLLS from this work (dot-dashed line, power-law index $\alpha = -1.19$ and $\alpha = -1.40$ for the upper and lower plots, respectively), and the low column density end of the double power-law fits to \hfreq\ for the DLA from PHW2005 (dashed line, power law index $alpha=-2.0$).  Neither fit describes the full range well, and the \hfreq\ shows a flattening near the canonical DLA definition of \lnhi$ = 20.3$\cmm.
}
\label{fig_fndla}
\end{figure}

Figure \ref{fig_fndla} shows the \hfreq\ distributions for the SLLS
and SDSS DLA to a column density of \lnhi $=21.5$\cmm.  
As before, we show the results for the different SLLS \nhi\ groups separately.  
Overplotted on the data are the SLLS single power--law fit and 
the low column density end of the double power-law for the SDSS-DLA (PHW2005). 
It is apparent that a simple extrapolation of the SDSS-DLA fit 
significantly overpredicts the frequency of the SLLS, especially
at \lnhi$< 19.7$ \cmm.
We contend, therefore, that the \hfreq\ flattens in slope around the \nhi\ corresponding to the canonical definition for the DLA, \lnhi $=20.3$\cmm. \cite{zheng02} predicted a flattening in \hfreq\ based on photoionization and self-shielding models of isothermal gas profiles in dark matter halos.  Although their analysis examined individual halos at a given mass, Zheng (2006, priv.\ comm.) has convolved their results with a Sheth-Tormen halo mass function \citep{Sheth99} and predict \hfreq\ should flatten at $\mnhi \approx 10^{20} \cm{-2}$, consistent with our results.

There are a few cautionary remarks to make regarding Figure~\ref{fig_fndla}.
First, the shape of the selection functions $g(z)$ for the SLLS and DLA samples
do not exactly match because the SLLS database includes a somewhat higher
fraction of $z>3$ quasars, and thus a fractionally larger $g(z)$ at those redshifts. Although the mean differs by only $\delta z = 0.3$,
we note that the comparison is not perfect as we have not considered any
evolution in the normalization of \hfreq\ for the SLLS or DLA but have simply
plotted the full samples.
Another systematic effect 
is that the SDSS-DLA sample may suffer from a Malmquist bias.
Specifically, the statistical and systematic errors 
(e.g. the effects of line blending) in the \nhi\ values of the DLA
are significant and will drive the observed \hfreq\ distribution to a steeper
slope.  It is possible that this effect explains the 
marked drop in \hfreq\ at $\mnhi = 10^{20.7} \cm{-2}$
in Figure~\ref{fig_fndla}.  We intend to address this issue directly
with follow-up, higher resolution observations of a large sample
of SDSS-DLA with $\mnhi \approx 10^{20.3} \cm{-2}$.
If there is a substantial Malmquist bias in the DLA sample, 
then the decrease in the slope of \hfreq\ would be more gradual than
that suggested by Figure~\ref{fig_fndla}.


\subsection{Is There an Inflection in \hfreq\ within the SLLS Range?}
\label{sec:inflect}

While the mismatch between the DLA and the \lya\ forest
in the power-law description of their \hfreq\ distributions
suggests that \hfreq\ for the LLS will 
show intermediate values ($\alpha \approx -1.7$),
the observed incidence of LLS reveals a different result.
As PHW2005 discussed, a simple spline interpolation of the DLA and
\lya\ forest \hfreq\ distributions through the LLS regimes predicts over
an order of magnitude more LLS than observed per $\Delta X$.
PHW2005 argued, therefore, that the \hfreq\ distribution for the LLS
must exhibit an inflection 
as evidenced by $d\log f/ d\log N > -1.5$.
\cite{zheng02} have also predicted that there should be an inflection
in \hfreq\ in the SLLS regime for galaxies exposed to an ionizing
radiation field.  It is worth investigating with our data-set whether
evidence exists for just such an inflection.

The simplest approach is to examine whether $d\log f/ d\log N > -1.5$
in the LLS regime.  Regarding our results on the SLLS, we find
that $d\log f/ d\log N > -1.5$ for both the $\log \mnhi^{lim} = 19.0$ 
\cmm and $\log \mnhi^{lim} = 19.3$ \cmm groups (Table~\ref{tab_fn}).
The differences, however, have less than $2\sigma$ significance.
Using only the current data-set and the distribution of DLAs, 
we do not report the existence of an inflection in \hfreq\ 
within the SLLS regime.  

\subsubsection{Constraints from lower \nhi\ LLS}
In order to pursue the question further, we introduce two new observational constraints on the LLS.   The statistical significance for the SLLS alone is limited
by the 
combination of sample size and observed baseline in \lnhi.  We cannot
arbitrarily increase the sample size, but we are able to introduce new 
constraints which are sensitive to lower column density LLS.
The number density of optically thick LLS has been well constrained by
many studies.  For our purposes here, we apply the constraint 
at redshift $z=3$ of $\ell(X)_{LLS} = 0.7 \pm 0.1$ \citep{ssb, 
storrie94, peroux_dla03}.  We also use a measure of the incidence
of optically thin partial Lyman Limit Systems (PLLS) from Burles (1997),
who found 12 systems with mean redshift at $z=3$ 
with 17.2 \cmm $< \mlnhi <$ 17.8 \cmm\ over a redshift
path of $\Delta z_{PLLS} = 59.07$ and absorption path of $\Delta X_{PLLS} = 16.63$. 
This gives $\ell(X)_{PLLS} = 0.20 \pm 0.06$. 

\subsubsection{The observable distribution function \O}
In order to facilitate the analysis and interpretation of the distribution
over the full range of LLS and DLAs, we introduce a function called the
observable distribution function of \hi:
\begin{equation}
{\mathcal O}(N,X) = \frac{m}{\Delta\log N \Delta X} \;
= \frac{\ell(X)}{\Delta\log N} \; ,
\end{equation}
where m is the number of systems observed over an absorption path, $\Delta X$,
and a column density range $\Delta\log N$.  
This is simply the frequency distribution
function in logarithmic column density bins and is related to the classic 
distribution by:
\begin{equation}
{\mathcal O}(N,X) \, d{\rm log } N = f(N,X) dN \, , \\ 
{\mathcal O}(N,X) = f(N,X) \, N \, \ln(10) \; .
\end{equation}

The observable distribution has a few nice features.  It is unitless, and
gives the direct number of systems observed over a specified bin in 
logarithmic column density (almost all studies of \ion{H}{1} absorption show and 
analyze the data in bins of constant width in \lnhi).  It also removes one
factor of column density from the steep slope in the frequency distribution, 
which enables better assessments of change in slopes, as well as smaller 
effects from the rapid change in the distribution over bins of large size.

\subsubsection{\O for the LLS and the DLA}
\begin{figure}
\epsscale{0.8}
\includegraphics[width=3.5in, trim= 0 0 0 0 ]{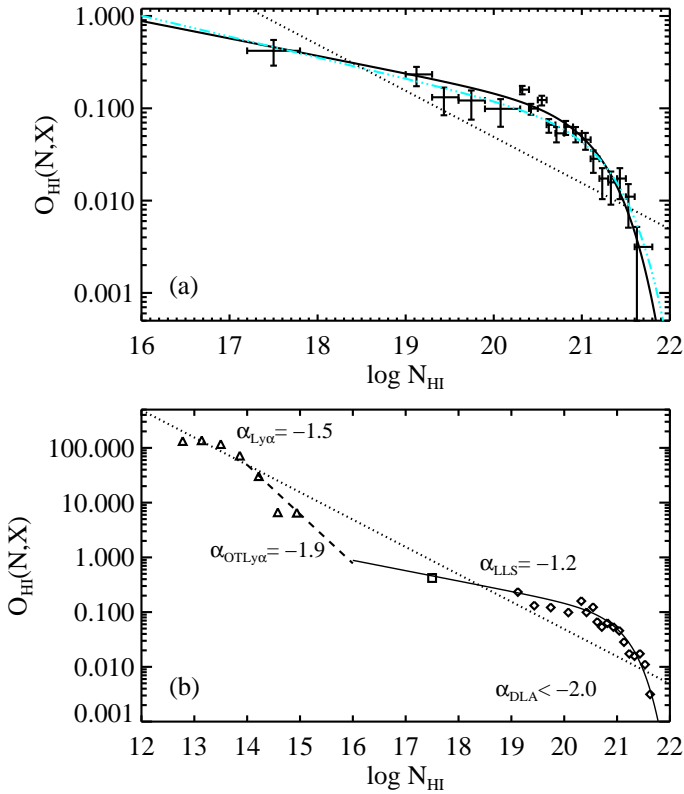}
\caption{a)  Values for the observable distribution  function for the entire range of H~I absorbers which are optically thick at the Lyman limit.  The solid line is a best fit 3 parameter function (assuming Gaussian statistics) of the data which include the SLLS presented here, the DLA from PHW2005, and a bin which contains the constraint on the number of partial LLS.  The dot-dashed line shows 
the best fit after removing the two outlying bins from the PHW sample.  
Also shown (dotted line) is the \O\ for the \lyaf.  
b) Same as a), but extended to 
include the \lyaf\ results of \citep{kt97}, which are shown as the triangles. The constraint from the number of partial LLS is displayed by the square. Overlaid are the values for the power-law slope $\alpha$ for 
the \nhi\ regions: $\alpha_{Ly\alpha}$ corresponding to the optically thin \lyaf\ with $12.1 < $ \lnhi\ $< 14$, $\alpha_{OTLy\alpha}$ corresponding to the optically thick \lyaf\ with $14 < $ \lnhi\ $< 16$, $\alpha_{LLS}$ corresponding to $16 <$ \lnhi $< 20.3$, and $\alpha_{DLA}$ corresponding to $\log \mnhi = 20.3 - 21.3$,
to highlight the observed and predicted changes in the logarithmic slope of \hfreq.
}
\label{O3_lls_DLA}
\end{figure}

In Figure~\ref{O3_lls_DLA}, we show the observational constraints on \O\ 
from the present sample of SLLS and the DLA from PHW2005.  In addition,
we show an observational constraint on the abundance of partial Lyman limits
at $z =3$, ${\mathcal O}_{PLLS} = 0.42 \pm 0.13$,
corresponding to the column density interval 
\citep[17.2 \cmm $< \mlnhi <$ 17.8 \cmm;][]{burles97}.  

We fit a simple analytic function to the combined set of ALL-19.0 SLLS
(Table~\ref{tab_fn}),
the PHW2005 DLA sample for \lnhi $>20.3$ \cmm, as well as the two additional
LLS constraints described above.
We fit a three parameter model expressed as:

\begin{equation}
{\rm log} \, {\mathcal O}(N,X) = a + b * (\mlnhi - 19) - 10^{\mlnhi - c}
\end{equation}
This parameterized form can be recast into a more familiar $\Gamma$-distribution
\citep[e.g.][]{pf95,peroux_dla03}:

\begin{equation}
{\mathcal O}(N,X) =  {\mathcal O_0} \,  
\left(\frac{\mnhi}{N_*}\right)^{\alpha+1} \, e^{-\frac{\mnhi}{N_*}}  
\end{equation}
where ${\rm log} \, N_* = c - 0.3622,  \; \;  
{\rm log} \, {\mathcal O_0} = b \, {\rm log} \, N_* + (a - 19 b), \; \;  
\alpha = b - 1 $.

We show the best fit function to the binned data assuming Gaussian
statistics as the solid line in Figure~\ref{O3_lls_DLA}, 
parameterized by $(a,b,c) = (-0.621, -0.189, 21.51)$, 
which gives a reduced $\chi^2$ for 17 degrees of freedom (dof), 
$\chi_\nu^2 = 1.86$.  We assess the dependence
of the fitting on the included data-sets, by sequentially removing constraints.
If we do not include the constraint on the total number of optically thick
Lyman limits, the best fit parameters are virtually unchanged, and the
fitted parameters above give a predicted number of optically thick Lyman limits of:
$\ell(X)^{fit}_{LLS} = 0.68$.
The two outliers in the above fit are the DLA points centered near bins of 
\lnhi = 20.3 and 20.6 \cmm.  If we drop the first of these points, 
we find $(-0.660, -0.198, 21.56)$ and $\chi_\nu^2 = 1.24$ for 15 dof.  
Dropping both DLA bins gives an acceptable fit, with 
$(-0.677, -0.227, 21.63)$ and $\chi_\nu^2 = 0.552$ for 14 dof.
The last fit is shown as a dashed-dotted line in Figure~\ref{O3_lls_DLA}.

The fits show that the \O\ slope in the SLLS region falls between 
$-0.227 < b < -0.189$ (recall $\alpha = b - 1$).
We also find that the HI cutoff scale is between 
$21.51 < c < 21.63$  which matches the results of PHW2005 
and \cite{peroux_dla03}.
All the fits produce a reasonable number of optically
thick Lyman limit systems: $0.61 < \mllls < 0.68$.

\subsubsection{Are the SLLS a distinct population?}

\begin{figure}
\epsscale{0.8}
\includegraphics[width=3.5in, trim= 0 0 0 0 ]{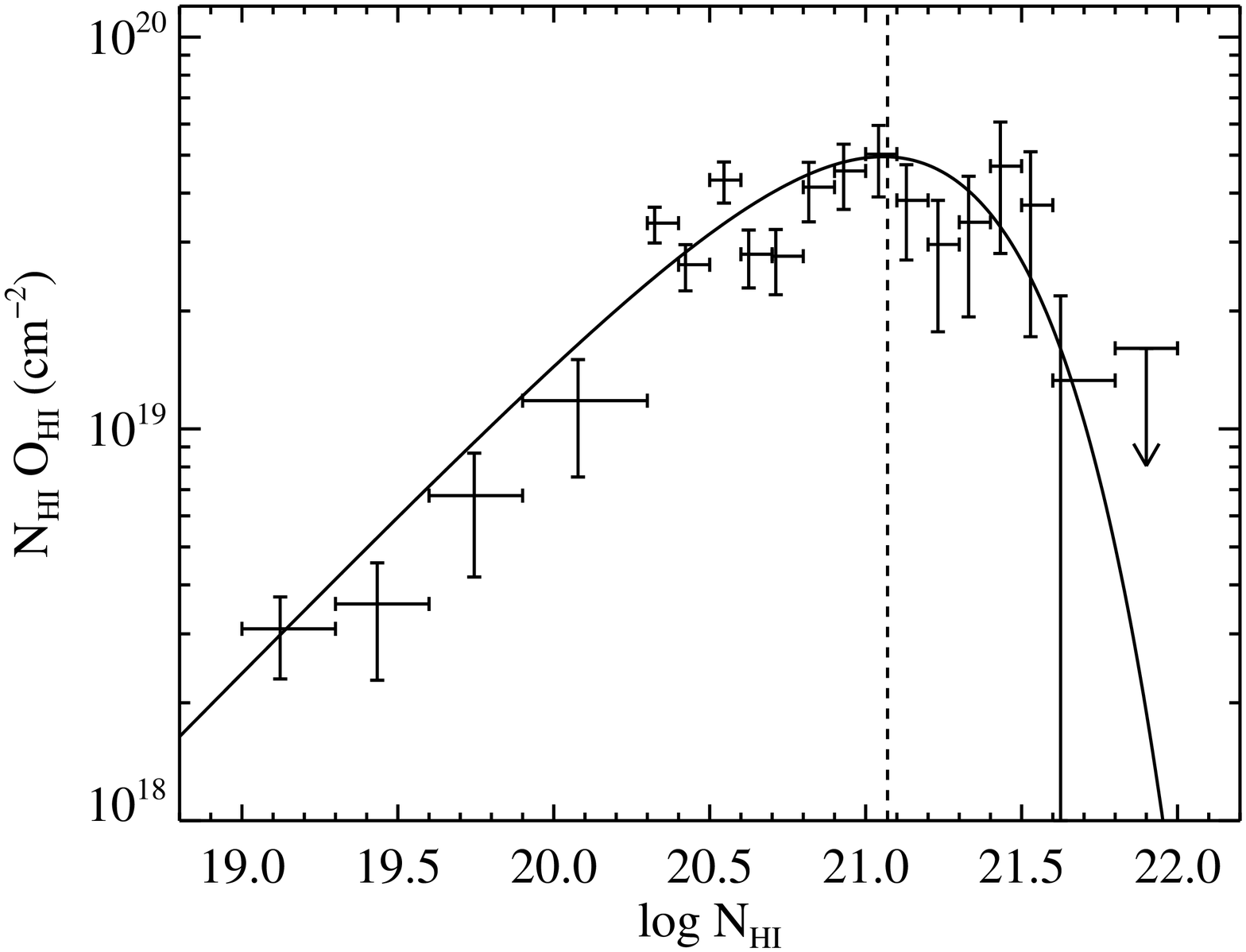}
\caption{Values for \nhi \O, the first moment of the observable 
distribution function over the SLLS and DLA range.  The data give the 
relative contributions to the cosmic density of hydrogen atoms per unit \lnhi\ 
at $z\sim 3$, and show that 
DLA with \lnhi = 20.3 -- 21.5 overwhelmingly dominate the \ion{H}{1} mass density, with the SLLS contributing only $\approx 15\%$ to the total.
}
\label{NO3_lls_DLA}
\end{figure}

Finally, we present another moment of the \nhi\ distribution, \nhi \O, in 
Figure~\ref{NO3_lls_DLA}. 
This represents the total \ion{H}{1} column density per unit logarithmic column
density per unit absorption path.  The SLLS and DLA data 
points and the first analytic fit
(to the entire data set) is shown.  
The data are relatively flat from $\mnhi = 10^{20.3} \cm{-2}$ to
$10^{21.5} \cm{-2}$ with the functional form peaking at 
$log \mnhi = 21.1$.
It is evident that absorbers with $\mnhi \approx 10^{21} \cm{-2}$
dominate the mass density of \ion{H}{1} in the universe.
Consider a comparison of the SLLS and DLA.
Whereas the damped \lya\ systems with $20.3 < \log \mnhi < 21.5$ dominate
the mass density with roughly equal contribution per $\Delta \log N$,
the SLLS lie distinctly below, with the absorbers at
$\mnhi = 10^{19}$\cmm\ adding a negligible portion and the full SLLS range in \nhi\ contributing only $\approx 15\%$ to the total mass density.  We note that
\cite{zwaan05} obtain a very similar shape for \nhi \O at $z=0$.
Again, this behavior is likely related to the fact that 
the majority of SLLS are highly ionized.
The results in Figure~\ref{NO3_lls_DLA}
lends further support to the concept that the SLLS absorbers 
are a distinct population from the damped \lya\ systems.

\subsection{Implications for lower \nhi}
Armed with a description for the full range of H~I column densities which are optically thick at the Lyman limit, we now wish to extend our analysis downward in \nhi\ to the \lyaf.
Although a more thorough analysis is required to fully assess the 
statistically acceptable distributions over such a large range in \lnhi,
we can clearly show that slopes in the SLLS region as steep as 
$\alpha = -1.5$ are unacceptable.  
In Figure~\ref{O3_lls_DLA}b, we overplot the extrapolation of the best fit 
observable distribution 
of low column density Lyman-$\alpha$ absorbers ($12.1 < \mlnhi < 14$) 
as a dotted line with a slope of $b = -0.5$ 
\citep[i.e., \hfreq$\propto N^{-1.5}$;][]{kt97}. 
Although the extrapolation has uncertainties related to the normalization and 
completeness, it does highlight the overprediction of SLLS based
on a simple power-law extrapolation from lower column density studies.
Even if we allowed for freedom in normalization, it is clear that in no region
except near \lnhi = 20.5 \cmm\ is $\alpha = -1.5$ a good description of the high
\lnhi\ distribution.  A subset of the data which were used constrain the 
fit to Lyman-$\alpha$ absorbers is shown in the lower panel of 
Figure~\ref{O3_lls_DLA} as triangles.\footnote{Note, the two 
points above \lnhi = 14.5 \cmm were not included in the power-law fit.}   
A gamma function with low-end slope of
$\alpha = -1.2$ is a good fit to the high density data sets of LLS and DLAs.
But the extrapolation of the low end slope to the regime of the 
Lyman-$\alpha$ absorbers underpredicts the observed numbers by almost a 
factor of 100!  

In order to reconcile the data presented here, together
with the DLAs, LLSs and the Lyman-$\alpha$ absorbers, the full distribution
function must contain at least three changes in logarithmic slope of the
frequency distribution (or inflections, $d^2 f/ d N^2 = 0$),
as previously argued (in part) by \cite{bechtold87} and \cite{petit93}.
The first inflection is seen in the change between the DLA distribution and SLLS.
This is required if one includes constraints from the LLS and the PLLS
and demands constant slopes down to these column densities.
There must then be at least two more changes in the logarithmic slopes, one to
account for the much higher number of Lyman-$\alpha$ absorbers, and the
second to finally merge back to the Lyman-$\alpha$ slope with $\alpha = -1.5$.
We present one such solution with a dotted line bridging the gap between 
\lnhi = 14 \cmm\ and \lnhi = 16 \cmm\ with a frequency slope of $\alpha = -1.9$.
The Lyman-$\alpha$ absorbers are claimed to have a single power law slope
over two decades of \nhi (12.5\cmm $< \mlnhi <$14.5\cmm), 
and are classified as a single population.  In comparison, 
the absorbers spanning 17\cmm $< \mlnhi <$20\cmm, over three decades of \nhi,
are well described by a single power law slope of $\alpha = -1.2$ and also 
could be classified as a single population by applying the same argument.

\section{\bf  Concluding Remarks} 

The SLLS results presented in this paper motivate several avenues of 
future exploration.  
First, we must extend the survey downward in LLS H~I column density.
An improved constraint on the number of optically thick LLSs would significantly tighten the constraint on the low-end slope of the SLLS distribution.  Second, a large sample of higher H~I column density \lyaf\ data is needed.  Specifically, a large sample of \lyaf\ with 
$14.5 \le$ \lnhi\  $\le 16.5$ \cmm\  directly tests
the predicted shape of the \hi distribution.  
Finally, one must pursue detailed ionization studies of the LLS
to fully assess their baryonic mass, metallicity, etc.
Our survey will
provide the data-set required for just such studies.

\acknowledgments

This paper includes data gathered with the 6.5 meter Magellan Telescopes 
located at Las Campanas Observatory, Chile.
The authors wish to recognize and acknowledge the very significant
cultural role and reverence that the summit of Mauna Kea has always
had within the indigenous Hawaiian community.  We are most fortunate
to have the opportunity to conduct observations from this mountain.
JO and SB acknowledge support from NSF grant AST-0307705.
GEP and JXP are supported by NSF grant AST-0307408.

\end{document}